\documentclass[aps,prd,twocolumn]{revtex4}

\usepackage{}
\usepackage{mathrsfs}
\usepackage{exscale}                 
\usepackage[intlimits]{amsmath}       
\usepackage{amsfonts}
\usepackage{amssymb,amscd}
\usepackage[dvips]{epsfig}
\usepackage{array}
\usepackage{wrapfig}
\usepackage{color}

\newcommand{\beqa}{\begin{eqnarray}}
\newcommand{\eeqa}{\end{eqnarray}}
\newcommand{\beq}{\begin{equation}}
\newcommand{\eeq}{\end{equation}}

\begin{document}
\title{A thermodynamically consistent quasi-particle model without density-dependent infinity of the vacuum zero point energy}
\author{Liu-jun Luo$^{1}$, Jing Cao$^{2}$, Yan Yan$^{2}$, Wei-Min Sun$^{2,3,4}$ and Hong-Shi Zong$^{2,3,4}$}
\address{$^{1}$ Key Laboratory of Modern Acoustics, MOE, Institute of Acoustics, and Department of Physics, Nanjing University, Nanjing 210093, P. R. China}
\address{$^{2}$ Department of Physics, Nanjing University, Nanjing 210093, China}
\address{$^{3}$ Joint Center for Particle, Nuclear Physics and Cosmology, Nanjing 210093, China}
\address{$^{4}$ State Key Laboratory of Theoretical Physics, Institute of Theoretical Physics, CAS, Beijing 100190, China}
\date{\today}

\begin{abstract}

In this paper, we generalize the improved quasi-particle model proposed in J. Cao et al., [ Phys. Lett. B {\bf711}, 65 (2012)] from finite temperature and zero chemical potential to the case of finite chemical potential and zero temperature, and calculate the equation of state (EOS) for (2+1) flavor Quantum Chromodynamics (QCD) at zero temperature and high density. We first calculate the partition function at finite temperature and chemical potential, then go to the limit $T=0$ and obtain the equation of state (EOS) for cold and dense QCD, which is important
for the study of neutron stars. Furthermore, we use this EOS to calculate the quark-number density, the energy density, the quark-number susceptibility and the speed of sound at zero temperature and finite chemical potential and compare our results with the corresponding ones in the existing literature.

\bigskip

E-mail: zonghs@chenwang.nju.edu.cn. ~~PACS numbers: 12.38.Mh, 51.30.+i, 52.25.Kn.

\end{abstract}

\maketitle

\section{Introduction}

The investigation of the equation of state (EOS) for cold and dense strongly interacting matter is still a contemporary focus \cite{1,2,3,4,5,6,61,62,7,72,71}. The EOS represents an important interrelation of state variables describing matter in thermodynamical equilibrium and we can get all the thermodynamic properties of the system from it. It is also well known that in astrophysics the study of the neutron star depends crucially on the EOS \cite{8,9,Tian,Yan}, because it provides the key information needed to evaluate stellar equilibrium configurations (e.g, recent astronomical observations have measured the highest neutron star mass ever determined in a precision weighing \cite{2solarmass}).

Nowadays Quantum Chromodynamics (QCD) is the generally accepted fundamental theory of strongly interacting matter. According to the phase diagram of QCD, at zero temperature and sufficiently high density, there is a phase transition from hadronic matter to a state which can be called plasma built of quark-gluon constituents (QGP). However, at present it is very hard to get a reliable EOS of cold and dense quark matter from the first principle of QCD. Lattice calculations are of limited use at finite chemical potential due to the appearance of the notorious sign problem. In addition, some analytical perturbative approaches have been considered, but in the physically relevant region, due to the large coupling, these methods seem basically to fail \cite{10}. In this case, one has to develop various phenomenological models which incorporate some basic features of QCD. Up to now, there have been many different QCD models \cite{11,12,13}. Among those models, the quasi-particle quark-gluon plasma model with few fitting parameters was widely used to describe the nonperturbative behavior of EOS observed in Monte Carlo (MC) simulations in the case of finite temperature \cite{14,15,16,17,18,19,33,34,35} (here it should be noted that due to lack of the data of MC simulations at finite chemical potential, there are very limited number of papers to apply quasi-particle model to study the EOS of QCD at finite density). In the quasi-particle model, at finite temperature, instead of real quarks and gluons with QCD interactions, one can consider the system to be made up of non-interacting quasi-particles with temperature-dependent effective mass, quasi-quarks and quasi-gluons. This model was first proposed by Goloviznin and Satz \cite{20}, and later by Peshier et al. \cite{21}. After some time, Gorenstein and Yang found that this model is thermodynamically inconsistent, and remedied this flaw by reformulating statistical mechanics \cite{22}. After that, Bannur also proposed a new quasi-particle model using standard statistical mechanics and avoided the thermodynamical inconsistency from the energy density rather than the pressure \cite{19}. Further, in Ref. \cite{23} Gardim and Steffens showed that the two models proposed by Peshier and Bannur are two extreme limits of a general formulation. Nevertheless, as the authors pointed out in Ref. \cite{24}, in all early works on quasi-particle model the problem of the temperature-dependent infinity of the vacuum zero point energy and its influences have not been seriously considered. In Ref. \cite{24}, the authors used the series expansion method inspired by Walecka \cite{25} to deal with the temperature-dependent infinity of the vacuum zero point energy and constructed a new thermodynamically consistent framework of quasi-particle model for QGP without the need of any reformulation of statistical mechanics or thermodynamical consistency relation. So, when Peshier et al. and Bannur generalized their respective quasi-particle models to the case of finite chemical potential, there will automatically emerge a term of chemical potential-dependent infinity of the vacuum zero point energy in the partition function, in other words, the partition function of quasi-particle model at finite chemical potential is also ill-defined.  In order to overcome this flaw, in the present paper we will generalize the improved  quasi-particle model proposed in Ref. \cite{24} from finite temperature and zero chemical potential to the case of zero temperature and finite chemical potential to make the partition function
of the quasi-particle model at finite chemical potential well-defined..

This article is organized as follows. In section II, we shortly review the problem existing in previous quasi-particle models for the purpose of self-consistency. In sections III, we generalize our improved quasi-particle model from finite temperature and zero chemical potential to the case of zero temperature and finite chemical potential. We first calculate the partition function of (2+1) flavor QGP at finite temperature $T$ and chemical potential $\mu$, then go to the limit $T=0$ in order to find the EOS at zero temperature. Furthermore, we calculate the quark-number density, the energy density, the quark-number susceptibility and the speed of sound of (2+1) flavor QGP at zero temperature and finite chemical potential and compare our results with the corresponding results in the existing literature. Finally we conclude our work with a summary.

\section{Problems in previous models}
We take the spin $1/2$ field as an example to illustrate the problem hidden in early quasi-particle models (similar problems also exist in the case of scalar field). As is well known, one always uses the Dirac field to characterize a system composed of Fermi-type quasi-particles. Now, let us begin with the Lagrangian of quasi-particle with a chemical potential-dependent mass $m=m(\mu)$
\begin{equation}
\mathscr{L}=\bar{\psi}(i\gamma^{\mu}\partial_{\mu}-m(\mu))\psi,
\end{equation}
where the effective mass $m(\mu)$ describes the interaction between real particles.

Inserting the Fourier expansion of the field $\psi(x)$ in imaginary time
\begin{displaymath}
\psi_{\alpha}(\vec{x},\tau)=\frac{1}{\sqrt{V}}\sum_{n}\sum_{\vec{p}}
\exp{[i(\vec{p}\cdot\vec{x}+\omega_{n}\cdot\tau)]\widetilde{\psi}_{\alpha;n}(\vec{p})}
\end{displaymath}
into this Lagrangian and according to the definition of partition function in the path integral formalism \cite{26,27}, one obtains
\begin{displaymath}
\begin{split}
Z&=\mathrm{Tr}e^{-\beta H}\\&=\int\limits_{antiperiodic} i \mathscr{D} \psi^{\dag}\mathscr{D}\psi\exp\bigg[\int\limits_{0}^{\beta}d\tau\int d^3x\mathscr{L}\bigg].
\end{split}
\end{displaymath}
Here, the above formula expresses the partition function $Z$ as a functional integral over $\psi$ of the exponential of the action in imaginary time. Then, using the standard path integral procedure, we can obtain the following partition function:
\begin{eqnarray}
\ln Z=2V\int\frac{d^3p}{(2\pi)^3}&\bigg[&\beta\omega^{\ast}+\ln(1+e^{-\beta(\omega^{\ast}-\mu)})\nonumber\\
&+&\ln(1+e^{-\beta(\omega^{\ast}+\mu)})\bigg],
\end{eqnarray}
where $\omega^{\ast}=\sqrt{p^{2}+m^{2}(\mu)}$ is the dispersion relation for the quasi-particle. From this expression we can see that the second and third part of the right-hand-side of Eq. (2) has the form of the ideal Fermi gas formula for a quasi-particle except for the chemical potential-dependent dispersion relation coming from the effective mass $m(\mu)$; the first term of the partition function is the zero point energy of the ``vacuum'' and is divergent as before, the difference between our case and that of standard statistical mechanics being that the infinity depends on the chemical potential $\mu$. In standard statistical mechanics we have $m=const$, and that divergence is independent of the chemical potential $\mu$, so we can throw this divergent part away in this case (this is because the vacuum zero-point energy and pressure cannot be measured experimentally and therefore the zero-point energy should be subtracted). This operation will have no effect on the computation of thermodynamical quantities. Thus, when we take the limit $T\rightarrow0$, we have:
\begin{eqnarray}
\mathcal {P}(T&=&0,\mu) =\lim_{T \rightarrow 0}\frac{1}{\beta}\frac{\partial\ln Z}{\partial\ V}\\
&=&\frac{4}{16\pi^{2}}\bigg[\frac{2}{3}\mu p^{3}_{F}-m^{2}\mu p_{F}+m^{4} \ln \bigg(\frac{\mu+p_{F}}{m}\bigg)\bigg],\nonumber
\end{eqnarray}
\begin{equation}
n(T=0,\mu)=\lim_{T \rightarrow 0}\frac{\partial \mathcal {P}(T,\mu)}{\partial \mu}=\frac{4 p_{F}^{3}}{6\pi^{2}},
\end{equation}
\begin{eqnarray}
\mathcal {E}(T&=&0,\mu)=-\mathcal {P}+\mu n\\
&=&\frac{4}{16\pi^{2}}\bigg[2\mu^{3} p_{F}-m^{2}\mu p_{F}-m^{4} \ln \bigg(\frac{\mu+p_{F}}{m}\bigg)\bigg],\nonumber
\end{eqnarray}
where $p_{F}=\sqrt{\mu^{2}-m^{2}}$ is the Fermi momentum. The above formula is formally model independent because it completely comes from ensemble theory. We can see that the divergent part vanishes automatically after taking the derivative, because it is only an infinite constant. Now let us turn back to our current effective Lagrangian. When the function $\omega^{\ast}(p)$ for the particle (``quasi-particle'' in our case) excitation energy becomes chemical potential-dependent, the operation of taking the derivative with respect to $\mu$ will no longer be valid. In particular, in the definition of number density, we find there is an extra term coming from the divergent part. If one wants to obtain some physically meaningful results, one must treat this infinity carefully rather than throws it away naively. Indeed, just as was shown above, the partition function is ill-defined. The next section of this paper is to generalize the improved quasi-particle model proposed in Ref. \cite{24} to the case of finite chemical potential and zero temperature to make the partition function well-defined.

\section{Results of calculation using new model and discussion}

In this section, we first do our calculation for Dirac field at finite temperature and chemical potential and then perform the limit $T\rightarrow0$ in order to find the EOS at zero temperature. As we have done in Ref. \cite{24}, we introduce a classical background field $B$ (it is allowed to depend on the temperature and chemical potential) into the Lagrangian of the quasi-particle system. Thus, for a system composed of Fermi-type quasi-particles, we have the following Lagrangian
\begin{equation}
\mathscr{L}=\bar{\psi}(i\gamma^{\mu}\partial_{\mu}-m(\mu))\psi+B.
\label{fermilag}
\end{equation}
Inspired by the approach of Walecka \cite{25}, we shall use a
series expansion method to separate the divergence term, and then choose an appropriate classical background field $B$ that satisfies the following condition
\begin{equation}
\frac{B}{m_{0}^{4}}=-\mathrm{Tr}\int\frac{d^4{\bar k}}{(2\pi)^4}i\sum\limits_{n=1}^{4}\frac{(\eta-1)^n}{n(\not\!{\bar k}-1)^n},
\label{bagconstant}
\end{equation}
where $m_{0}$ is the rest mass of quasi-particle at zero chemical potential, $\eta=\frac{m(\mu)}{m_{0}}$ and ${\bar k}^\mu=\frac{k^\mu}{m_0}$ is the dimensionless four-momentum measured in unit of $m_0$, as the counterterm to remove the divergence and make the shift in the ground-state energy of the total system finite.

After taking into account the effect of this classical background field, by means of dimensional regularization we can eliminate the divergence of vacuum zero point energy and obtain the finite, physically meaningful result of the shift in the ground-state energy
\begin{eqnarray}
\Delta\varepsilon_{0}&&=E_{0}-B-E_{vac}\\
&&=-\frac{2}{(4\pi)^2}\bigg[m^{4}(\mu)\ln\frac{m(\mu)}{m_{0}}\nonumber\\
&&+m_{0}^{3}(m_{0}-m(\mu))-\frac{7}{2}m_{0}^{2}(m_{0}-m(\mu))^{2}\nonumber\\
&&+\frac{13}{3}m_{0}(m_{0}-m(\mu))^{3}-\frac{25}{12}(m_{0}-m(\mu))^{4}\bigg].\nonumber
\end{eqnarray}

Just as is shown above, by means of the series expansion method invented by Walecka we have eliminated the chemical potential-dependent infinity of the vacuum zero point energy. Here it should be noted that in the above calculation we have only considered one component of the vacuum energy-momentum tensor $T^{00}$. We also note that the counterterm contribution to the energy-momentum tensor that is produced by $B$ is proportional to the Minkowski metric. In the following we will prove that the vacuum energy-momentum tensor can also be rendered finite by introducing the classical background field.

According to the Lagrangian given in Eq. (6), one can derive the following energy-momentum tensor
\begin{equation}
T^{\mu \nu}=\bar{\psi}i\gamma^{\mu}\partial^{\nu}\psi-g^{\mu \nu}B. \nonumber
\end{equation}
By means of the Green function
\begin{eqnarray}
G(y,x)&=&<|\hat{T} \psi(y)\overline{\psi}(x)|> \nonumber
\\&=&\int\frac{d^4k}
{(2\pi)^4}\frac{i}{k\!\!\!/-m+i\epsilon}e^{-ik(y-x)}, \nonumber
\end{eqnarray}
we can express the vacuum energy-momentum tensor as
\begin{eqnarray}
T^{\mu \nu}&=&<|i\bar{\psi}\gamma^{\mu}\partial^{\nu}\psi|>-g^{\mu \nu}B\nonumber \\
&=&\lim_{y^{0}\rightarrow x^{0}}\lim_{\vec{y}\rightarrow \vec{x}}-\mathrm{Tr}i\gamma^{\mu}\partial_{y}^{\nu}G(y,x)-g^{\mu \nu}B
\nonumber \\
&=&-\mathrm{Tr} \int\frac{d^4k}{(2\pi)^4}i\gamma^{\mu}k^{\nu}\frac{1}{\not\!k-m+i\epsilon}-g^{\mu \nu}B.\nonumber
\end{eqnarray}
What we are interested in is the difference between the expectation value of the vacuum and that of the true ground state. Thus the measurable quantity in the experiment is
\begin{eqnarray}
&&\frac{<T^{\mu \nu}>-<T^{\mu \nu}>_{vac}}{m_{0}^{4}}  \\
&=&-\mathrm{Tr}\int\frac{d^4{\bar k}}{(2\pi)^4}i\gamma^{\mu}{\bar k}^{\nu}\left[\frac{1}{\not\!{\bar k}-\eta}-
\frac{1}{\not\!{\bar k}-1}\right]-g^{\mu \nu}\frac{B}{{m_0}^4}\nonumber \\
&=&-\mathrm{Tr}\int\frac{d^4{\bar k}}{(2\pi)^4}i\gamma^{\mu}{\bar k}^{\nu}\frac{\eta-1}{(\not\!{\bar k}-1)^2}
\left[\frac{1}{1-\frac{\eta-1}{\not\!{\bar k}-1}}\right]-g^{\mu \nu}\frac{B}{{m_0}^4}\nonumber \\
&=&-\mathrm{Tr}\int\frac{d^4{\bar k}}{(2\pi)^4}
i\gamma^{\mu}{\bar k}^{\nu}\sum\limits_{n=1}^{\infty}\frac{(\eta-1)^{n}}{(\not\!{\bar k}-1)^{n+1}}-g^{\mu \nu}\frac{B}{{m_0}^4}.\nonumber
\label{prove}
\end{eqnarray}
 Using the cyclic property of the trace operation, we can show that
\begin{displaymath}
\begin{split}
\mathrm{Tr}\int&\frac{d^4{\bar k}}{(2\pi)^{4}}i\gamma^{\mu}{\bar k}^{\nu}\sum\limits_{n=1}^{\infty}\frac{(\eta-1)^{n}}{(\not\!{\bar k}-1)^{n+1}}
\\&=-\mathrm{Tr}\int\frac{d^4{\bar k}}{(2\pi)^{4}}i{\bar k}^{\nu}\frac{\partial }{\partial {\bar k}_{\mu}}\sum\limits_{n=1}^{\infty}\frac{(\eta-1)^{n}}{n({\bar k}\!\!\!/-1)^{n}}.
\end{split}
\end{displaymath}
A partial integration with respect to ${\bar k}^{\mu}$ now reduces the integral in Eq. (9) to the form:
\begin{eqnarray}
&&\frac{<T^{\mu \nu}>-<T^{\mu \nu}>_{vac}}{m_{0}^{4}} \nonumber
\\
&=&-{\mathrm Tr}\int\frac{d^4{\bar k}}{(2\pi)^4}
ig^{\mu \nu}\sum\limits_{n=1}^{\infty}\frac{(\eta-1)^n}{n(\not\!{\bar k}-1)^n}-g^{\mu \nu}\frac{B}{m_0^4}. 
\label{Tmunudiff}
\end{eqnarray}
Now from Eq. (\ref{bagconstant}) and Eq. (\ref{Tmunudiff}), 
it can be seen that the first four divergent terms in the summation on the right-hand-side of Eq. (\ref{Tmunudiff}) are exactly canceled by the classical background field $B$, while according to the power-counting analysis, the remaining terms with $n\geqslant5$ have enough powers of $k$ downstairs to ensure convergence in the physical case of four dimensions. Therefore, by introducing one classical background field $B$ the shift in the ground-state energy-momentum of the total system can be rendered finite. This is what one expects in advance. As we have stressed in Ref. \cite{24}, the series expansion method inspired by Walecka's approach for eliminating the chemical potential-dependent infinity of vacuum zero point energy given in Eq. (\ref{bagconstant}) is the simplest way to achieve this goal.

After completing the elimination of the infinity of vacuum zero point energy successfully, it is easy to see that the partition function for the Dirac field is given by
\begin{eqnarray}
\ln Z&=&2V\int\frac{d^3p}{(2\pi)^3}\bigg[\ln(1+e^{-\beta(\omega^{\ast}-\mu)})\nonumber\\
&+&\ln(1+e^{-\beta(\omega^{\ast}+\mu)})\bigg]-V\beta\Delta\varepsilon_{0}.
\end{eqnarray}
Therefore, the pressure is
\begin{eqnarray}
\mathcal {P}_{q}=\frac{1}{\beta}\frac{\partial\ln Z}{\partial V}&=&2T\int\frac{d^3p}{(2\pi)^3}\bigg[\ln(1+e^{-\beta(\omega^{\ast}-\mu)})\nonumber\\
&+&\ln(1+e^{-\beta(\omega^{\ast}+\mu)})\bigg]-\Delta\varepsilon_{0},\nonumber
\end{eqnarray}
here we take the $T=0$ limit and obtain the EOS at zero temperature and finite chemical potential
\begin{eqnarray}
\mathcal {P}_{q}(T=0,\mu)&=&\frac{4}{16\pi^{2}}\bigg[\frac{2}{3}\mu p^{3}_{F}-m^{2}(\mu)\mu p_{F}\\
&+&m^{4}(\mu) \ln \bigg(\frac{\mu+p_{F}}{m(\mu)}\bigg)\bigg]-\Delta\varepsilon_{0},\nonumber
\end{eqnarray}
where $p_{F}=\sqrt{\mu^{2}-m^{2}(\mu)}$ is Fermi momentum. All states up to the energy $E_{F}=\mu$ are occupied and all states above it are empty.

The analysis of the divergence of vacuum zero point energy related to massless gauge bosons will be similar to the  previous one. We repeat the above analysis and arrive at the form of the classical background field for gauge bosons:
\begin{displaymath}
B=N\int\frac{d^4k}{(2\pi)^4}\frac{i}{2}\sum\limits_{n=1}^{2}(-1)^{n+1}\frac{m^{2n}(\mu)}{n(k^{2}-m^{2}(\mu))^{n}},
\end{displaymath}
where $N$ is the degeneracy factor ($N=6$ for SU(2); and $N=16$ for SU(3)). Here we still adopt dimensional regularization to calculate $E-E_{vac}$ and $B$. After some algebra we find:
\begin{equation}
\Delta\varepsilon_{0}=E_{0}-E_{vac}-B=\frac{N}{(4\pi)^2}\frac{1}{8}m^{4}(\mu).
\end{equation}
Therefore, in the case of gauge bosons, there is only one term proportional to the fourth power of effective mass contributing to the finite shift in the ground-state energy. The pressure is
\begin{equation}
\mathcal {P}_{g}=\frac{1}{\beta}\frac{\partial\ln Z}{\partial V}=-T\int\frac{d^3p}{(2\pi)^3}\ln(1-e^{-\beta\omega^{\ast}})-\Delta\varepsilon_{0}.\nonumber
\end{equation}
After taking the $T=0$ limit, the above equation can be reduced to the form
\begin{equation}
\mathcal {P}_{g}(T=0,\mu)=-\frac{N}{(4\pi)^2}\frac{1}{8}m^{4}(\mu).
\end{equation}
In summary, for a system composed of both Bose-type quasi-particles and Fermi-type quasi-particles at zero temperature and finite chemical potential, the pressure can be described by the following equation
\begin{eqnarray}
\mathcal {P}(0,\mu)=\mathcal {P}_{q}+\mathcal {P}_{g},
\end{eqnarray}
where $\mathcal {P}_{q}$, $\mathcal {P}_{g}$ is given by Eq. (12) and
Eq. (14), respectively. Quarks come in different flavors and also carry a color charge, thus these expressions need to be multiplied by $N_{f}N_{c}$ to take into account of that. According to statistical mechanics, once the EOS of the system is obtained, all the thermodynamical functions can be determined by fundamental thermodynamical relations. Here, it should be noted that since the starting point of our discussion is the partition function, all thermodynamical quantities in our model automatically satisfy the thermodynamical consistency relation. Therefore, we do not need to make use of any extra conditions to guarantee the thermodynamical consistency of our quasi-particle model, which is quite different from other quasi-particle models in the literature.

We apply our model to the case of (2+1) flavor QCD at zero temperature and sufficiently high density. Considering the interacting plasma in thermodynamic equilibrium, we assume that it can be described as a system of massive quasi-particles, a picture arising asymptotically from the in-medium properties of the constituents of the plasma. According to
Refs. \cite{29,30}, we write the effective mass of the quasi-particle as
\begin{displaymath}
m_{i}^{2}(T)=m_{0 i}^{2}+\Pi_{i},
\end{displaymath}
where $m_{0 i}$ and $\Pi_{i}$ are the rest mass and thermal mass of the quasi-particle, respectively.
$\Pi_{i}$ are given by the asymptotic values of the gauge-independent hard-thermal/density-loop self-energies:
\begin{displaymath}
\Pi_{q}=2\omega_{q 0}(m_{0}+\omega_{q 0}),
\end{displaymath}
\begin{displaymath}
\omega_{q 0}^{2}=\frac{N_{c}^{2}-1}{16N_{c}}\frac{\mu^{2}_{q}}{\pi^{2}}G^{2}(\mu),
\end{displaymath}
\begin{displaymath}
\Pi_{g}=\frac{1}{4\pi^{2}}\sum_{q}\mu^{2}_{q}G^{2}(\mu),
\end{displaymath}
\begin{displaymath}
G^{2}(T=0,\mu)=\frac{48\pi^{2}}{(11N_{c}-2N_{f})\ln(\frac{\mu+T_{s}}{T_{c}\pi/\lambda})^{2}},
\end{displaymath}
where $N_{c}$, $N_{f}$ stands for the color factor and the number of flavors, respectively, and $m_{0}$ is the rest mass of the quark. $T_{c}/\lambda$ is related to the QCD scale $\Lambda_{QCD}$. The quantity $G^{2}$ is to be considered as an effective coupling constant. Using the parameters $\lambda=6.6$, $T_{s}=-0.78 T_{c}$ and $T_{c}=170$ MeV
\cite{29,30}, we calculate the pressure density $\mathcal {P}$. It is interesting to compare the EOS obtained in this paper to the EOS of QCD proposed in previous studies. Here we shall take one prominent example, the cold, perturbative EOS of QCD proposed by Fraga, Pisarski, and Schaffner-Bielich in Ref. \cite{31}. The pressure density to second order in $\alpha_{s}$ in the $\overline{MS}$ scheme obtained in Ref. \cite{31} is quoted as follows:
\begin{eqnarray}
\mathcal {P}_{FPS}(\mu)&=&\frac{N_{f}\mu^{4}}{4\pi^{2}}\bigg\{1-2\bigg(\frac{\alpha_{s}}{\pi}\bigg)-\bigg[G+N_{f}\ln\frac{\alpha_{s}}{\pi}\nonumber\\
&+&\bigg(11-\frac{2}{3}N_{f}\bigg)\ln\frac{\overline{\Lambda}}{\mu}\bigg]\bigg(\frac{\alpha_{s}}{\pi}\bigg)^2\bigg\},
\end{eqnarray}
where $G=G_{0}-0.536N_{f}+N_{f}\ln N_{f}$, $G_{0}=10.374\pm0.13$ and $\overline{\Lambda}$ is the renormalization subtraction point. The scale dependence of the strong coupling $\alpha_{s}(\overline{\Lambda})$ is taken as
\begin{eqnarray}
\alpha_{s}(\overline{\Lambda})=&&\frac{4\pi}{\beta_{0}u}\bigg[1-\frac{2\beta_{1}}{\beta_{0}^{2}}\frac{\ln(u)}{u}+\frac{4\beta_{1}^{2}}{\beta_{0}^{4}u^{2}}\bigg(\bigg(
\ln(u)-\frac{1}{2}\bigg)^{2}\nonumber\\
&&+\frac{\beta_{2}\beta_{0}}{8\beta_{1}^{2}}-\frac{5}{4}\bigg)\bigg],\nonumber
\end{eqnarray}
where $u=\ln(\overline{\Lambda}^{2}/\Lambda_{\overline{MS}}^{2})$, $\beta_{0}=11-2N_{f}/3$, $\beta_{1}=51-19N_{f}/3$, and $\beta_{2}=2857-5033N_{f}/9+325N_{f}^{2}/27$.
For $N_{f}=3$, $\Lambda_{\overline{MS}}=365$ MeV. The only freedom in the model of Ref. \cite{31} is the choice of the ratio $\overline{\Lambda}/\mu$, which is taken to be 2 in that reference. A plot of our EOS and the EOS of Fraga, Pisarski, and Schaffner-Bielich is given in Fig. 1. It can be seen that in the region of $\mu$ studied, the pressure density in the EOS of Fraga, Pisarski, and Schaffner-Bielich and our EOSs increases monotonically as $\mu$ increases. In the large $\mu$ region the EOS of Fraga, Pisarski, and Schaffner-Bielich and our EOSs show qualitatively similar behaviors: all of them tend to the free quark gas result as $\mu$ tends to infinity. However, compared with the EOS of Fraga, Pisarski, and Schaffner-Bielich, our EOS tends somewhat more slowly to the free quark gas result in the region of $\mu$ studied. This shows that even in relatively large region of chemical potential, QGP is still strongly interacting.

Then let us investigate the quark-number density
\begin{equation}
n(\mu)=\frac{\partial \mathcal {P}}{\partial\mu}.\nonumber
\end{equation}
The dependence of $n(\mu)$ on $\mu$ is plotted in Fig. 2. We note that when $\mu$ is smaller than a critical value $\mu_{0}$, the quark-number density vanishes. This comes from $p_{F}>0$, so we have $\mu^{2}\geqslant m^{2}(\mu)$ and $\mu_{0}=260$ MeV. Namely, $\mu=\mu_{0}$ is a singularity which separates two regions with different quark-number densities. This result agrees qualitatively with the general conclusion of Ref. \cite{32}. In that reference, based on a universal argument, it is pointed out that the existence of some singularity at the point $\mu=\mu_{0}$ and $T=0$ is a robust and model-independent prediction. The numerical value of the critical chemical potential in pure QCD (i.e., with electromagnetic interactions being switched off) is estimated to $(m_{N}-16 MeV)/N_{c}=307$ MeV (where $m_{N}$ is the nucleon mass and $N_{c}=3$ is the number of colors). The value of the critical chemical potential obtained in this study is almost of the same order of magnitude as the estimate in that reference. While in Ref. \cite{31}, according to $\mathcal {P}_{FPS}=0$, those authors pointed out that the scope of application of the Eq. (16) is $\mu>425$ MeV. 

Just as shown in Fig. 2, the obtained quark-number density distribution differs significantly from the Fermi distribution of the free quark theory. Physically this is a consequence of dynamical chiral symmetry breaking in the low energy region. Now let us give an explanation for this point. It is well known that the quasi-particle model phenomenologically incorporates the effect of dynamical chiral symmetry breaking through its $T$- and $\mu$-dependent effective quasi-particle mass. This can be seen from the fact that the effective mass of the quasi-particle is much larger than the corresponding current quark mass.  This leads to a dynamical mass of the quasi-particle which is about $300~\rm{MeV}$. As we know, the nucleon is made up of three constituent quarks and the nucleon mass is about $900~\rm{MeV}$.  Hence, naively speaking, to excite a nucleon from the vacuum, one needs to provide an energy of about $300~\rm{MeV}$ per each quark (more detailed argument for this point can be found in Ref. [41]).
This is the physical reason why the critical chemical potential $\mu_0$ is about $300~\rm{MeV}$.

We also calculate the energy density
\begin{eqnarray}
\mathcal {E}=-\mathcal {P}+\frac{\partial \mathcal {P}}{\partial\mu}\mu \nonumber
\end{eqnarray}
and the quark-number susceptibility (QNS) \cite{He}
\begin{equation}
\chi=\frac{\partial n(\mu)}{\partial \mu}=\frac{\partial^{2}\mathcal {P}}{\partial \mu^{2}}.\nonumber
\end{equation}
The results for the energy density and QNS as a function of chemical potential are shown in Fig. 3 and Fig. 4. In Fig. 3 we can see a clear rise in the energy density at a chemical potential of about 400$\thicksim$500 MeV. As a comparison we also plot the energy density of Fraga {\it et al.}. Just as shown in Fig. 3, when $\mu$ is small, the result of Fraga {\it et al.} deviates from ours. It can be seen that when $\mu$
is below about $500~\rm{MeV}$, the energy density in the model of Fraga {\it et al.} is very large. As $\mu$ increases, the energy density drops suddenly at first and then increases slowly. 
This behavior is hard to understand. We think that this may implies that the EOS of Fraga 
{\it et al.} is applicable only for $\mu>500~\rm{MeV}$. Physically, dynamical chiral symmetry breaking is a nonperturbative phenomena and it cannot be treated using perturbative methods. Since dynamical chiral symmetry breaking plays an important role at low chemical potential, the EOS of Fraga {\it et al.} obtained using perturbative methods is not valid at low chemical potential. Only in the large $\mu$ region do both of the two results tend to the ideal quark gas result. Then, let us see the curve of $\chi$. One sees clearly from Fig. 4 that $\chi$ equals zero below a critical chemical potential $\mu_{0}=260$ MeV and experiences a sudden increase across it; when $\mu>\mu_{0}$, $\chi(\mu)$ increases monotonously with $\mu$. These characteristic of QNS is in qualitative agreement with the previous results \cite{36}. From Fig. 4 one can also see that when the chemical potential becomes large the QNS tends to the ideal quark gas result. Because of asymptotic freedom this tendency is what one expects in advance.

The speed of sound is also an important physical quantity, because it is related to the speed of small perturbations produced in QCD matter. On the other hand, if we want to study the transport properties of the plasma phase at zero temperature, we must know
\begin{displaymath}
C^{2}_{s}(T=0,\mu)=\frac{d\mathcal {P}}{d\mathcal {E}}.
\end{displaymath}
The curve of the speed of sound squared as a function of chemical potential $\mu$ is shown in Fig. 5. As can be seen in Fig. 5, as $\mu$ increases, the speed of sound squared tends to the ideal gas limit $(C^{SB}_{s})^{2}=1/3$. On the other hand, as in the previous case of energy density, here the speed of sound of Fraga et al. also shows unphysical effect in small chemical potential regions. In contrast, our result is more reasonable.

\begin{figure}[t!]
\centerline{\includegraphics[width=0.9\columnwidth]{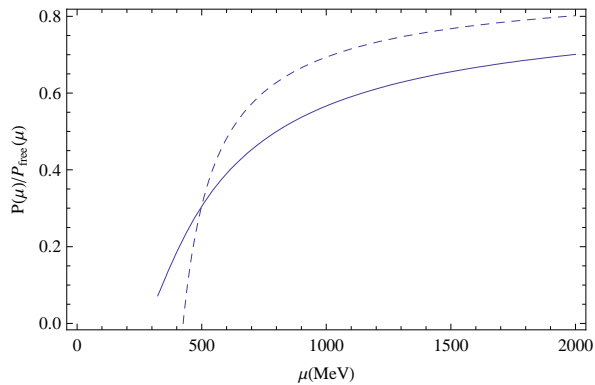}}
\caption{The pressure, normalized by the free quark gas pressure $\mathcal {P}=N_{c}N_{f}\mu^{4}/(12\pi^{2})$, in our EOS (15) (solid line) and the EOS (16) of Fraga, Pisarski, and Schaffner-Bielich (dashed line).}
\label{P}
\end{figure}

\begin{figure}[t!]
\centerline{\includegraphics[width=0.9\columnwidth]{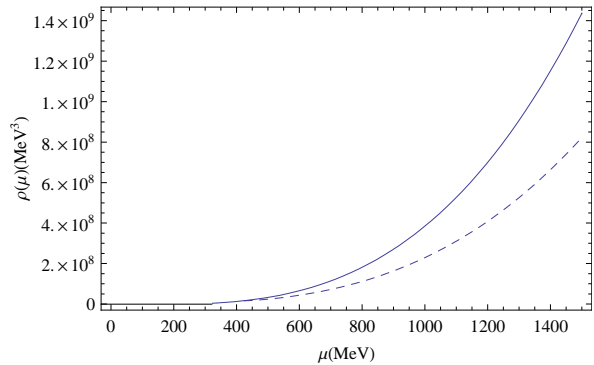}}
\caption{Relation between the quark-number density and the chemical potential $\mu$. The solid line represents our result and the dashed line represents the corresponding one of Fraga et al..}
\label{P}
\end{figure}

\begin{figure}[t!]
\centerline{\includegraphics[width=0.9\columnwidth]{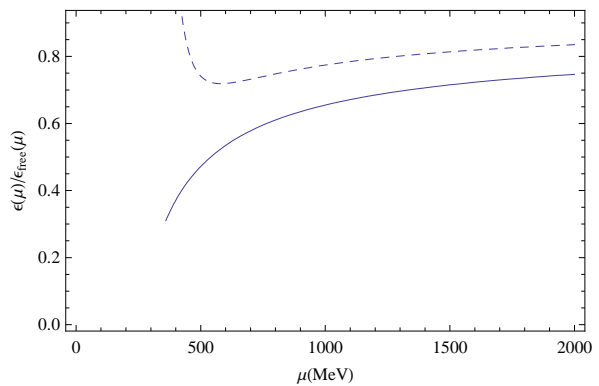}}
\caption{The comparison between the energy density $\mathcal {E}/\mathcal {E}_{free}$ obtained from our model of (2+1) flavor QCD plasma (solid line) and that obtained from Fraga et al.(dashed line). Here $\mathcal {E}_{free}$ is the energy density for an ideal quark gas in the chiral limit.}
\label{P}
\end{figure}

\begin{figure}[t!]
\centerline{\includegraphics[width=0.9\columnwidth]{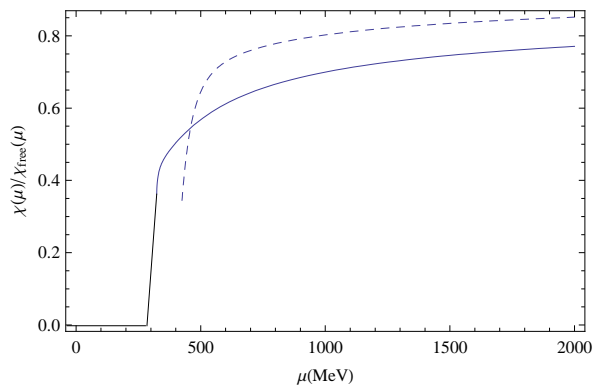}}
\caption{The comparison between the $\chi/\chi_{free}$ obtained from our model of (2+1) flavor QCD plasma (solid line) and that obtained from Fraga et al.(dashed line). Here $\chi_{free}$ is the QNS for an ideal quark gas in the chiral limit.}
\label{P}
\end{figure}

\begin{figure}[t!]
\centerline{\includegraphics[width=0.9\columnwidth]{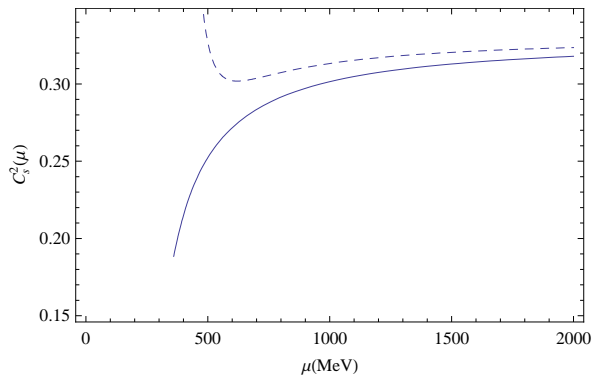}}
\caption{The comparison between the speed of sound squared in the (2+1) flavor QCD plasma obtained from our model (solid line) and that obtained from Fraga et al.(dashed line).}
\label{P}
\end{figure}

Now let us discuss the limitations and range of validity of our model. One usually thinks that as a phenomenological model, the quasi-particle model is valid at large chemical potential. However, one is not certain in which range of chemical potential this model is valid. The range of validity of the quasi-particle model can only be known from comparison of the results of this model with other theoretical and experimental results. From Fig. 1 and Fig. 2, it can be seen that in the large $\mu$
limit the quasi-particle model has the feature of asymptotic freedom, and when $\mu$ is below a critical value $\mu_0$, the quark number density in this model vanishes. Both two features are requirements of QCD and they are fulfilled in our model. Therefore, we think that our model may be valid when $\mu$ is larger than this critical chemical potential $\mu_0$.

Here it is also interesting to compare our model with the other quasi-particle models in the literature, e.g. the model by Peshier  {\it et al.}. Since the EOS plays an important role in the study of neutron star, we use the EOS as an example in our comparison to the results of the model by Peshier {\it et al.}.

In our model, the EOS is
\begin{eqnarray}
\mathcal {P}(\mu)&=&P_{id}-\Delta\varepsilon_{0},
\end{eqnarray}
where
\begin{eqnarray}
\Delta\varepsilon_{0}&=&E_{0}-B-E_{vac}=-\frac{2}{(4\pi)^2}\bigg[m^{4}(\mu)\ln\frac{m(\mu)}{m_{0}}\nonumber\\
&&+m_{0}^{3}(m_{0}-m(\mu))-\frac{7}{2}m_{0}^{2}(m_{0}-m(\mu))^{2}\nonumber\\
&&+\frac{13}{3}m_{0}(m_{0}-m(\mu))^{3}-\frac{25}{12}(m_{0}-m(\mu))^{4}\bigg].\nonumber
\end{eqnarray}
In the model of Peshier {\it et al.}, the EOS is
\begin{eqnarray}
\mathcal {P}(\mu)&=&P_{id}-B,
\end{eqnarray}
where
\begin{eqnarray}
B=B_{0}+\int_{\mu_{0}}^{\mu}\frac{\partial P_{id}}{\partial m}\frac{d m}{d \mu}d \mu\nonumber
\end{eqnarray}
Comparing Eqs. (17) and (18), it can be seen that in the above two EOS, the first term $P_{id}$, which represents the standard ideal gas expression with the effective mass $m(\mu)$, is the same. The only difference between these two EOS comes from the second term.
Numerical calculation shows that the difference between these two terms is not obvious. This is because in our work we use Peshier's ansatz of the effective mass of the quasi-particle and the parameters. Since numerically the difference between these two EOS is not obvious, we do not plot the curve of the EOS of the model of Peshier {\it et al.} separately.

Here it is interesting to compare the thermodynamical consistency of our model and the previous one. The authors of Ref. [32] introduced an additional requirement
 \begin{eqnarray}
 \bigg(\frac{\partial\mathcal {P}}{\partial c_{1}}\bigg)_{T,\mu,c_{2},\cdots}=0,
~~~~ \bigg(\frac{\partial\mathcal {P}}{\partial c_{2}}\bigg)_{T,\mu,c_{1},\cdots}=0,~~ \cdots
 \end{eqnarray}
to guarantee the thermodynamical consistency of quasi-particle model for $H=H_{eff}$ with $T$- and/or $\mu$-dependent parameters $c_{1},c_{2},\cdots$. In that paper they pointed out that these conditions were only introduced as self-consistency equations for the phenomenological extension of the mean-field theory. Here, it should be noted that since the starting point of our discussion is the partition function, all thermodynamical quantities in our model automatically satisfy the thermodynamical consistency relation. Therefore, we do not need to make use of these extra conditions to constrain the parameters of our model, which is quite different from other quasi-particle models in the literature.
\section{Summary and conclusions}

To summarize, in this paper the improved  quasi-particle model proposed in Ref. \cite{24} is generalized from finite temperature and zero chemical potential to the case of zero temperature and finite chemical potential in order to calculate the EOS of (2+1) flavor QCD at finite chemical potential and zero temperature. We first calculate the partition function at finite temperature and chemical potential, then go to the limit $T=0$ and obtain the EOS for cold and dense QCD. Furthermore, we use this EOS to calculate the quark-number density, the energy density, the quark-number susceptibility and the speed of sound. We can see that as the chemical potential increases, the behaviors of the pressure, the energy density, the quark-number susceptibility and the speed of sound tend to the free quark gas result. A comparison between our EOS and one prominent example of the EOS of QCD, the cold, perturbative EOS of QCD proposed by Fraga, Pisarski, and Schaffner-Bielich in Ref. \cite{31} is made. The quark-number density differs significantly from the Fermi distribution of free quark theory. Physically this is a consequence of dynamical chiral symmetry breaking in the low energy region. It is found that when $\mu$ is below a critical value $\mu_{0}$, the quark-number density vanishes. This feature agrees with the general conclusion in
Ref. \cite{32}. The value $\mu_{0}$ obtained here is almost of the same order of magnitude as the estimate made in Ref. \cite{32}. From the comparison between our results of energy density and the speed of sound and the corresponding ones of Fraga et al., we can see that when $\mu$ is smaller than 500 MeV, our model is more valid. Finally, we would like to stress that the lack of MC data on dense strongly interacting matter at zero temperature leads to the uncertainty of phenomenological parameters of quasi-particle model at zero temperature and finite chemical potential.
In our future work, we hope to use the most recent astronomical observations \cite{2solarmass} to constrain the parameters of the quasi-particle model at finite chemical potential.

\section{acknowledgments}

This work is supported in part by the National Natural Science Foundation of China (under Grant 11275097, 10935001, 11274166 and 11075075), the National Basic Research Program of China (under Grant 2012CB921504) and the Research Fund for the Doctoral Program of Higher Education (under Grant No 2012009111002).

\end{document}